\documentstyle[draft,epsf,12pt]{article}  
\begin{document}  
\centerline{\normalsize\bf   
ON THE RESONANT SPIN FLAVOR PRECESSION OF THE NEUTRINO }  
\baselineskip=16pt  
\centerline{\normalsize\bf IN THE SUN}  
\vspace*{2cm}
\setcounter{footnote}{1} 
\renewcommand{\thefootnote}{\fnsymbol{footnote}}
\centerline{\footnotesize  
J. Derkaoui and Y. Tayalati\footnote{Also at CEA/Saclay F-91191 
Gif/Yvette Cedex, France. E-Mail: tayalati@hep.saclay.cea.fr}}  
\renewcommand{\thefootnote}{\arabic{footnote}}
\baselineskip=13pt  
\centerline{\footnotesize\it  
University Mohamed I$^{st}$, Faculty of Sciences}  
\centerline{\footnotesize\it Dept. of Physics, LPTP. BP. 524}  
\centerline{\footnotesize\it 60000 Oujda, Morocco}  
\centerline{\footnotesize\tt  
derkaoui@sciences.univ-oujda.ac.ma}  
\centerline{\footnotesize\tt  
tayalati@sciences.univ-oujda.ac.ma}  
\normalsize\baselineskip=15pt  
\vspace*{1cm}  
\centerline{\today}  
\vspace*{3cm}  
\setcounter{footnote}{0}
			\begin{abstract}  

	This work deals with the possible solution of the solar neutrino
	problem in the framework of the resonant neutrino spin-flavor 
	precession scenario.        
	The event rate results from the  solar neutrino experiments as well as
 	the recoil electron energy spectrum from SuperKamiokande are used to 
constrain the free parameters of the neutrino in this model ($\Delta m^2$ and 
$\mu_\nu$) . We consider two kinds of magnetic profiles inside the sun. For 
both cases, a static and a twisting field are discussed.
			\end{abstract}

\section{Introduction}  
	
	The amount of accurate solar neutrino data available at present, the
	numerous cheks of the functioning of the solar neutrino detectors that
	have been and are being performed together with more precise results in
	the field of solar modeling which are in a very impressive agreement with
	 high-accuracy helioseismological data, suggest strongly that the	
	observed deficiency of the solar neutrino is one of the most
	convincing indications for new physics beyond the standard electroweak
	theory.
	
	Two attractive solutions to this puzzle are (i) the Mikheyev 
	- Smirnov -   Wolfenstein~(MSW)  \cite{msw}  
	matter-enhanced neutrino oscillations, and (ii) the spin precession via  
	a large magnetic moment of the electron neutrino($\mu_\nu$) proposed by 
	Okun, Voloshin and  Vysotsky~(OVV)\cite{ovv} motivated by the apparent 
	anticorrelation of neutrino flux  in the Davis experiment with the sunspot 
	activity\cite{sun}. This precession is enhanced  resonantly in the presence of
 matter	yelding to a resonant spin flavor  precession~(RSFP)\cite{rsfp} 
	analogous to the MSW effect.\\  
	The present paper deals with the RSFP type solution to the solar  
        neutrino problem. This solution has been advocated by  many 
	authors[5,6].
The magnitude and the profile of the sun magnetic field beeing the most 
uncertain variables, they have been assigned a variety of values and forms 
(for a review see [5,6]).It is worth noticing that the solar neutrino 
experimental results may be used to probe the solar interior 
(its magnetic field profile for instance) \cite{pulido2}.
We explore the preliminary 708~days\footnote{The present statistics of SuperKamiokande
is about 825 days.} 
	of operation results from  the SuperKamiokande~[8,9] experiment 
        together with the last results of the four solar neutrino  
        experiments (Homestake\cite{exp1}, Kamiokande\cite{exp5},  
        SAGE\cite{exp3} and GALLEX\cite{exp2}) for two  
	variants of the solar magnetic field. 
	The range for the RSFP neutrino parameters $\Delta m^2$ and $\mu_\nu$
	compatible with the total measured event rates in all of the solar neutrino
	experiments, as well as with the SuperKamiokande electron recoil energy
	spectrum are investigated both for a static and a twisting magnetic field.
	The paper is organised as follows:  
        In section 2 a brief review of the experimental situation is presented.  
        In section 3 the neutrino propagation equation through solar matter is  
        solved analytically using the Landau Zener\cite{lz} approximation. 
	Hence we get the neutrino parameters allowed by the total event rates in
	neutrino experiments and the distortion of the recoil electron energy
	spectrum measured by SuperKamiokande.
	The evolution equation and the allowed regions for  
        neutrino parameters are next considered in section 4 taking into account  
        the twisting effect of the sun magnetic field. Our conclusions  
        are summarised in section 5. Some details about the contour finding  
	procedure are presented in the appendix.

\section{Experimental status}  
        Let us first briefly recall the experimental situation. Table \ref{data}  
        summarize the neutrino event rates that have been measured in  
        the four pioneering solar neutrino experiments Homestake\cite{exp1} ,  
        Kamiokande\cite{exp5},  
        SAGE\cite{exp3} and GALLEX\cite{exp2} together with the $^8B$ neutrino flux measured after  
        708 runing days by SuperKamiokande[8,9]. The observed event rates are  
  significantly smaller than the theoretical expectation of the BP98 model\cite{bp98}.\\   
        The GALLEX and SAGE experiments measure the same quantity, in what follows  
        we consider only their weighted average rate. We also adopt the  
        SuperKamiokande measurement as the most precise direct determination of the  
        higher energy $^8B$ neutrino flux.  
\section{Static field}  
\subsection{Time evolution of the solar neutrino}  
        Disregarding, for simplicity, possible neutrino mixing,  
        the time evolution equation of the neutrino in the transverse magnetic  
        field B is expressed in the Majorana\footnote{For Dirac neutrino, just replace $\frac{5}{3}$ in the   
$G_F.n_e$ coefficient by $\frac{11}{6}$.} weak  
        interaction ($\nu_{{e}}$,$\bar\nu_{{\mu}}$) sector as:  
\begin{equation}  
i\frac{d}{dt}  
\left [  
\begin{array}{c}  
\nu_{e}\\  
\bar\nu_{\mu}  
\end{array}  
\right ]  
=  
\left [  
\begin{array}{cc}  
\frac{5}{3}\frac{G_F}{\sqrt{2}}n_e  & \mu B \\  
\mu B & \frac{\Delta m^2}{2E}  
\end{array}  
\right ]  
\left [  
\begin{array}{c}  
\nu_{e}\\  
\bar\nu_{\mu}  
\end{array}  
\right ],  
\end{equation}  
where $\Delta m^2$ is the neutrino flavour mass square difference, and   
$G_F$  
is the Fermi coupling constant. We used the approximation $n_n \simeq \frac{1}{6}n_e$ between the electron  
and neutron densities in the sun, valid in the convection and upper radiation zones where most of the neutrino trajectory  
lies\cite{n6e}.\\ The electron density $n_e$ decreases exponentially along the   
neutrino trajectory and is well approximated by\cite{n6e}.  
\begin{equation}  
G_F.n_e = 2.11 \times 10^{-11}\exp{(-\frac{r}{0.09R_\odot})} \;\;\; eV,  
\end{equation}  
r being the distance from the center of the sun and $R_\odot$ being the  
solar radius.\\  
We use the analysis of Parke\cite{parke} to determine the average   
$\nu_{e}$ survival probability:   
\begin{equation}  
P_{\nu_{e}\rightarrow\nu_e}   
=   
\frac{1}{2}+(\frac{1}{2}-P_{LZ})\cos{2\hat{\theta}_i}\cos{2\hat{\theta}_f},  
\label{p}  
\end{equation}  
where $\hat{\theta}_{i(f)}$ denotes the initial(final) value of the mixing  
angle $\hat{\theta}$ such that,  
  
\begin{equation}  
\sin^2{\hat{\theta}}  
=   
\frac{\mu^2B^2}  
{\mu^2B^2  
+\frac{1}{4}  
\left \{   
\frac{5G_F}{3\sqrt{2}}n_e - \frac{\Delta m^2}{2E}  
+ \sqrt{(  
\frac{5G_F}{3\sqrt{2}}n_e - \frac{\Delta m^2}{2E}  
)^2  
+ 4\mu^2B^2} \right \}^2 },  
\end{equation}  
and the jump probability $P_{LZ}$ is taken in the Landau~Zener\footnote{The Landau-Zener  
approximation which assumes a linearly decreasing density in the vicinity of  
the critical point works rather well in the sun\cite{pulido}}  
approximation\cite{lz}  
\begin{equation}  
P_{LZ} =  
\exp  
\left [  
- 2 \pi \frac{(\mu B)^2}{\frac{\Delta m^2}{2E}\frac{1}{n_e}  
|\frac{dn_e}{dr}|}  
\right ],  
\label{lz}  
\end{equation}  
calculated  at the resonance  point obtained by requiring,  
\begin{equation}  
\frac{5G_F}{3\sqrt{2}}n_e - \frac{\Delta m^2}{2E} = 0.  
\label{res}  
\end{equation}  

 	The generalisation of eq.3 to include the neutrino energy and production range
	distributions is thus:

\begin{equation}
P_{\nu_e\rightarrow\nu_e} = \int w_E w_r (\frac{1}{2}+
(\frac{1}{2}-P_{LZ}))\cos{2\hat{\theta}_i}\cos{2\hat{\theta}_f} dE dr_i,
\end{equation}
	where the function $w_E$, $w_r$ represent respectively the
	probability density of neutrino production per unit energy and per unit
	length.

	Unfortunately, the magnetic field inside the sun is not accessible to direct
	observation. At the moment there is no model for the solar magnetic field
	and very little is known about it: not only
	its profile is unknown, but even its strenght is very uncertain.\\
	An upper limit on the strengh of the solar magnetic field comes
	from the requirement that the field pressure must be smaller than that of matter.
	This is a rather weak limit (for the convective zone $B<10^7G$) and all
	other more stringent bounds are highly model dependent. Several plausible
	profiles have been proposed and investigated in literature. In the following
	we consider two distributions of the solar magnetic field:

\begin{enumerate}  
\item Linear distribution:   
\begin{equation}  
\left \{  
\begin{array}{ll}  
B = 10^5 G & r \leq 0.7 R_{\odot}   \\  
B = 10^5 G(1-3.33(r/R_\odot - 0.7))& r > 0.7 R_{\odot}.     
\end{array}  
\right .  
\label{bl}  
\end{equation}  
\item Wood-Saxon (WS) distribution:   
\begin{equation}  
B(r) = \frac{10^5G}{1 + \exp{10(r - R_\odot)/R_\odot}}.  
\label{bws}  
\end{equation}  
\end{enumerate}  
\subsection{Total event rates}

Using only the total event rates measured at the Homestake, Gallium and
Super\-Kamiokande\footnote{Since the quoted uncertainty in the Kamiokande rate
is much larger than the uncertainty in the SuperKamiokande, the results are
essentially unchanged if the rate from kamiokande is also considered. }
experiments, figure 1(2) shows the allowed region in the
($\mu_\nu,\Delta m^2$) parameters space for the linear (Wood-Saxon) profile.
The black dot within each allowed region indicates the position of the best
fit point in the parameters space. 
The best fit for the linear profile is obtained for:
\begin{equation}  
\begin{array}{ll}  
\Delta m^2 & = 1.8 \times 10^{-8}eV^2,   \\  
\,\,\mu & = 3.9 \times 10^{-12} \mu_B,  
\end{array}  
\end{equation}  
for which $\chi^2_{min} = 1.57$.  
 For the Wood-Saxon profile, the best fit occurs at:
\begin{equation}  
\begin{array}{ll}  
\Delta m^2 & = 1.7 \times 10^{-8}eV^2,   \\  
\,\,\mu & = 7.7 \times 10^{-11} \mu_B,  
\end{array}  
\end{equation}  
with $\chi^2_{min} = 0.94$.

\subsection{Recoil electron energy spectrum}

	Unlike the neutrino event rate deficit, the energy spectrum 
	of recoil electron observed at SuperKamiokande is one of the most 
	important model independent solar observables. 
	Therefore, a deviation of the observed electron recoil energy 
	spectrum shape from what is predicted by standard electroweak theory would 
	be an indication of new physics (such as neutrino oscillations).

	The SuperKamiokande experiment\cite{exp4} has 
	measured the energy spectrum of recoil electrons from the  
	neutrino-electron  elastic scattering in water above 5.5~MeV. In what follows we
will use the SuperKamiokande data relative to 708 days of operation.
The data are given in an 18-bin energy histogram, where
for each bin we have  the ratio between the experimental event rate and the
theoretical one. The first 17 bins have a width of 0.5~MeV starting from 5.5~MeV 
while the last bin include events with energies from 14~MeV to 20~MeV.

Fig.3 and fig.4 show (respectively for linear and W-S magnetic field
profiles) the RSFP neutrino parameters regions that are allowed when
we take into account the information from the SuperKamiokande recoil electron
energy spectrum alone. For both profiles, a large region of RSFP parameters space is
consistent with the  data. The best fit is obtained with the linear
magnetic field at:

\begin{equation}  
\begin{array}{ll}  
\Delta m^2 & = 3.0 \times 10^{-8}eV^2,   \\  
\,\,\mu & = 3.1 \times 10^{-12} \mu_B,  
\end{array}  
\end{equation}  
with $\chi^2_{min} = 0.90$.\\
For the Wood-Saxon profile the best fit occurs at:
\begin{equation}  
\begin{array}{ll}  
\Delta m^2 & = 1.0 \times 10^{-8}eV^2,   \\  
\,\,\mu & = 3.7 \times 10^{-11} \mu_B,  
\end{array}  
\end{equation}  
and $\chi^2_{min} = 1.08$.

Fig. 5 and fig.6 show the regions compatible with both the event rates and  
the information from SuperKamiokande spectrum data.
Even though the best fit solution considering only the spectrum information
(the dark point in figures 3 and 4) do not lie within
the allowed regions by the analysis taking into account 
the event rate information, a large region in the neutrino parameters space
($\Delta m^2,\frac{\mu}{\mu_B}$)
are consistent with constraints from the total measured event rates as well
as with those  from the SuperKamiokande recoil electron energy spectrum.
In fig.7 we plot the recoil electron energy distribution divided by the standard
prediction expected to be observed in the SuperKamiokande detector, using the
best fit parameters found (eqs.10 and 11 respectively for the linear and
the W-S profiles) together with the data from SuperKamiokande\cite{exp4}.    

\section{Twisting magnetic field}  
  
        Several works proposed that the transverse component of  
        the solar magnetic field may change its direction along the neutrino  
        trajectory\cite{rot}. This can lead to new interesting phenomena in neutrino  
        physics\cite{new}. In this section we report on the observed deficit   
        interpreted in terms of RSFP for a twisting magnetic field. We also
	discuss how such rotation would affect the recoil electron energy spectrum 
	observed by the SuperKamiokande experiment. 
  
	The equation for the flavor neutrino wave function describing the  
	propagation of neutrino in matter with twisting transverse magnetic   
	field can be written -for the case of interest- as:  
\begin{equation}  
i\frac{d}{dt}  
\left [  
\begin{array}{c}  
\nu_{e}\\  
\bar\nu_{\mu}  
\end{array}  
\right ]  
=  
\left [  
\begin{array}{cc}  
\frac{5}{3}\frac{G_F}{\sqrt{2}}n_e + \dot{\phi}  & \mu B_T \\  
\mu B_T & \frac{\Delta m^2}{2E}  
\end{array}  
\right ]  
\left [  
\begin{array}{c}  
\nu_{e}\\  
\bar\nu_{\mu}  
\end{array}  
\right ],  
\end{equation}  
	The angle $\phi(t)$ defines the direction of the magnetic field  
$\vec{B_T}(t)$ in the orthogonal plane to the neutrino momentum and   
$B_T=|\vec{B_T}(t)|$. \\  
The evolution equation looks like the one obtained in the static  
case (no twisting) with the addition of a quantity proportional to 
$\dot\phi$ to the effective matter density . The formalism sketched above  
for the calculus of the neutrino survival probability remains valid.  
  
	Let us assume that $\dot\phi \sim \frac{1}{r_0}$ where $r_0$  
is the curvature radius of the magnetic field lines, then it is easy to  
see that the effect of the twisting field becomes significant when:  
\begin{equation}  
\frac{1}{r_0} \sim \frac{5}{3}\frac{G_F}{\sqrt{2}}n_e.  
\end{equation}  
    For matter density values taken at the bottom of the  
    convective zone of the sun, this gives:  
\begin{equation}  
\label{order}  
		  r_0 \sim 0.1 R_\odot.  
\end{equation}  
         
We did the same calculation as before assuming  
both linear and W-S profiles for $B_T(r)$ and taking for  
$k=R_\odot/r_0$ (a dimensionless factor which characterise  
the twisting velocity)  
the values $\pm 10$ according to the interesting feature expected  
from eq.\ref{order}.\\  
        Fig.8 shows the allowed region obtained using the linear profile  
while fig.9 gives the analogous result for the W-S profile.\\  
        For k=+10, the  
two field configurations used in our calculations give poor fits to  
the total event rates. ($\chi^2_{min} = 8.77$ and $\chi^2_{min} = 8.15$ for the W-S  
and the linear profiles respectively) \\  
        In contrast, when the field twists in the opposite side (k=-10), good fits  
to the total event rates are obtained with both configurations of the magnetic
field.  
  
        Using the linear field distribution, the best fit is  
 found for:  
\begin{equation}  
\begin{array}{ll}  
\Delta m^2 & = 9.9 \times 10^{-11}eV^2,      \\  
\,\,\mu & = 2.3 \times 10^{-12} \mu_B,  
\end{array}  
\end{equation}  
with a shallow $\chi^2_{min} = 0.18$.
In an analogous way, we find  the best fit to the data for the W-S field at:  
\begin{equation}  
\begin{array}{ll}  
\Delta m^2 & = 8.4 \times 10^{-11}eV^2,      \\  
\,\,\mu & = 3.4 \times 10^{-11} \mu_B, 
\end{array}  
\end{equation}  
with $\chi^2_{min} = 0.2$.
         
The most important change in the allowed range of neutrino parameters  
compared to the standard case~(no twisting) is the disappearence of the area at
the lower right corner of figs.1-2 compatible with the measured experimental event rates at
the 99\%CL. The  allowd regions are also extended to somewhat smaller values of 
$\Delta m^2$. The main reason  
for this is the fact that the resonant density depends on the magnitude of  
$\Delta m^2$ as well as on the velocity $\dot\phi$ which has a tendency to  
move it inward the sun, so for $\Delta m^2$ small enough, different solar  
 neutrinos types can also undergo resonant  
transition in both convective and radiative zones.  
 
 Fig.10 and fig.11 show the regions of neutrino parameters compatible with the 
 SuperKamiokande spectrum of the recoil electron energy for the field distributions 
given by (8) and (9) and the two twisting magnitudes (k=$\pm$10).\\ 
In the case k=10, better fits to the SuperKamiokande
 electron spectrum  are obtained compared to the total event rates case. 
For a linear distribution we have: 

\begin{equation}  
\begin{array}{ll}  
\Delta m^2 & = 1.0 \times 10^{-7}eV^2,      \\  
\,\,\mu & = 2.9 \times 10^{-12} \mu_B,\\  
\,\,\chi^2_{min} & = 0.68,  
\end{array}  
\end{equation}  

while for a W-S distribution, the result  is:
 
\begin{equation}  
\begin{array}{ll}  
\Delta m^2 & = 8.0 \times 10^{-8}eV^2,      \\  
\,\,\mu & = 3.8 \times 10^{-11} \mu_B,\\  
\,\,\chi^2_{min} & = 0.81,  
\end{array}  
\end{equation}  

Good fits are also obtained for the case (k=-10) and they give:
 
\begin{equation}  
\begin{array}{ll}  
\Delta m^2 & = 7.6 \times 10^{-9}eV^2,      \\  
\,\,\mu & = 2.7 \times 10^{-12} \mu_B,\\  
\,\,\chi^2_{min} & = 1.45,  
\end{array}  
\end{equation}  
 
 for the linear field and:
 \begin{equation}  
\begin{array}{ll}  
\Delta m^2 & = 3.1 \times 10^{-9}eV^2,      \\  
\,\,\mu & = 4.3 \times 10^{-11} \mu_B,\\  
\,\,\chi^2_{min} & = 1.52,  
\end{array}  
\end{equation}  
 for the W-S field.
              
	      This results can be easily understood if 
 we consider  the dependence of neutrino survival
 probability on the twisting magnitude $\dot\phi$ (see fig.12). 
 For a given $\Delta m^2$,  at energies of the solar neutrino spectrum low enough,
  $\dot\phi$ is very small with respect to 
 the neutrino oscillation coeficient $\frac{\Delta m^2}{2E}$.
 It follows that the twisting effect is absent at this energy scale.
 The twisting effect  
        becomes significant for the highest energy part of the spectrum and depends on the  
        sign of $\dot\phi$. RSFP is amplified when (k=-10) and suppressed  
        for the opposite sign. This dependence  is in favour of a positive twisting~(k=10)
	to fit the electron spectrum measured by the SuperKamiokande  
        experiment.

Comparing the regions allowed by the total event rates shown in figs.8 and 9 with 
those allowed by the energy spectrum shown in
Figs.10 and 11, it follows (Fig.13-14) that for k=+10, none of the field
distribution is able to explain the whole data from underground 
experiments. For the case k=-10, there is a large region in ($\Delta m^2$,
$\frac{\mu}{\mu_B}$) compatible with both constraints for the two fields. For this
later case, and using the best fit parameters found (eqs.21-22), we plot in 
figure~15 the expected RSFP distortion of the SuperKamiokande recoil energy
spectrum.

We stress that we have also studied two extreme cases corresponding to  
slow ($k\leq 1$) and fast($k\geq 100$) twisting. While no apreciable change  
from the standard case (no twisting) has been found for the case of  
slow twisting, the highest values of k decouples the   
($\nu_e,\bar{\nu_\mu}$)system and consequently the fast twisting scenario  
is surely unable to explain the deficit found by the underground  
experiments. 
\section{Conclusion:}  

The results from the SuperKamiokande experiment opens a new area in the solar neutrino
studies. Using the Landau-Zener formalism, we have investigated the RSFP way for
a solution to the solar puzzle in the light of the latest experimental results as
well as the theoretical predictions.\\
We have identified the allowed regions of neutrino parameters for either static
and twisting magnetic field in the sun. This was done first by considering the constraints 
from total event rate results. The obtained subset of the parameters that are
consistent with the total rates was then confronted to the  SuperKamiokande
electron recoil energy spectrum to extract  possible neutrino parameter regions
compatible with the whole data set.
We found that the RSFP scenario can account for both the observed deficiency of
solar neutrino flux and the measured SuperKamiokande spectrum. This is the case
of the static field and of the rotating one with negative k. Although it provides
a good fit to the spectrum data, solar twisting field with positive k seems to be disfavoured to
explain the whole data available from underground experiments.\\
We should stress that the quality of the fits depends also on the chosen field configuration.
Differences between the two cases as large as one order of magnitude in the 
$\mu$ values are found.   \\
We emphasize that the somewhat vagueness of the  
conclusions drawn by this work are related to the poor knowledge we have  
of the field distribution and -in the case of a rotating field-  of the magnitude  
and the sign of the rotation velocity $\dot\phi$.\\  
It is crucial to have a good enough  knowledge of the sun magnetic field   
parameters in order to draw an ultimate conclusion to the RSFP contribution  
to the solution of the solar neutrino puzzle.\\
It is also of great importance to compare the RSFP with other oscillation 
scenarios(i.e. the MSW, the just so and the vacuum solutions) to set the more
likely solution. The present energy spectrum data are uncertain enough(especialy
in hight energy part of the spectrum) to inhibit any clear statement. However,
the growing experimental accuracy will - in a near future-  allow to validate/exclude  more 
easily the various proposed models. 
The possible time variation (anticorrelation with the Wolf cycle of the sun activity) 
 that may be observed in the SK signal should also help making the comparison.

\section*{Appendix} 
Given the experimental rates $R_i$ and their uncertainties $\sigma_i$, the $\chi^2$ is 
defined as:
\begin{equation}
\chi^{2}_{rates}=\sum_{i,j}[R_{i,th}-R_{i,exp}] 
[\sigma^{2}_{ij,tot}]^{-1}[R_{j,th}-R_{j,exp}],
\end{equation}
where i,j label the experiment type (Ga, Cl and SK). The expected rates and the error matrix 
are derivated following the method of \cite{lisi} and
\begin{equation}
\sigma^{2}_{ij,tot}=\sigma^{2}_{ij,exp}+\sigma^{2}_{ij,th},
\end{equation}
\begin{equation}
\sigma^{2}_{ij,exp}=\delta_{ij}\sigma_{i,exp}\sigma_{j,exp},
\end{equation}
denoting by $\sigma_{i,exp}$, the experimental error given in table 1.\\
$\sigma^{2}_{ij,th}$ is the sum of two contributions: the one coming from uncertainties on 
cross sections (CS) and the one from the uncertainties on astrophysical parameters (AP)
\begin{equation}
\sigma^{2}_{ij,th}=\sigma^{2}_{ij,CS}+\sigma^{2}_{ij,AP}.
\end{equation}
If we write a particular rate event as:$R_{i}=\sum C_{ij}\phi_{j}$ \\
then we have
\begin{equation}
\sigma^{2}_{ij,CS}=\delta_{ij}\sum_{k,l=1}^{8}\frac {\partial R_i}{\partial lnC_{kj}} 
\frac{\partial R_j}{\partial lnC_{lj}}\Delta lnC_{kj}\Delta lnC_{lj}\\
=\delta_{ij}\sum_{k=1}^{8} (R_{ik}\Delta lnC_{ik})^2, 
\end{equation}
where $R_{ik}=C_{ik}\phi_k$ are the partial rates and the sum runs over the eight relevant
neutrino fluxes (i.e.$ pp,pep,hep,^{7}Be,^{8}B,^{13}N,^{15}O$ and $^{17}F$).\\ 
Similarly we obtain:
\begin{equation}
\sigma^{2}_{ij,AP}= =\delta_{ij}\sum_{k,l=1}^{8} 
R_{ik}R_{jl}\Delta ln\phi_{k}\Delta ln\phi_{l}.
\end{equation}
The $\Delta ln\phi_{i}$ are calculated using the Bahcall's code exportrates.f \cite{code} with
some minor modifications.\\
The correlation matrix used in this work is:
\begin{center}
\begin{tabular}{lccc}
\hline
\hline
Experiment & & Correlation matrix & \\
\hline
\hline
Ga. & 1.000 & & \\
Cl. & 0.671 & 1.000 & \\
SuperKamiokande & 0.687 & 0.964 & 1.000 \\
\hline
\hline
\end{tabular}
\end{center}

On the other hand, for the study of the observed SuperKamiokande energy spectrum, we use 
the following $\chi^2$:
\begin{equation}
\chi^{2}_{spec}=\sum_{i,j}[\beta S_{i,th}-S_{i,exp}] 
[W^{2}_{ij}]^{-1}[\beta S_{j,th}-S_{j,exp}],
\end{equation}
where $S_{i,th}$ is the predicted event rate for the i-th energy bin, $S_{i,exp}$ is the 
measured rate and $\beta$ is a  free parameter which normalizes the predicted $^{8}B$ 
solar neutrino flux to the measured flux.  

The input neutrino fluxes have been taken  from\cite{bp98}.  
Asymmetric errors have been conservatively  
symmetrized to the largest one. We use the improved neutrino cross section  
for each detector given in  \cite{cros} and the neutrino spectra  
given in  \cite{spec}.   
Allowed regions in the ($\frac{\mu}{\mu_B}$,$\Delta m^2$) plan are obtained  
by finding the minimum $\chi^2$ and plotting contours of constant  
$\chi^2 = \chi^2_{min} + \Delta\chi^2$ where $\Delta\chi^2=4.61$ for 90\% C.L.,  
5.99 for 95\% C.L. and 9.21 for 99\% C.L. 
All results are given as function of the reduced 
$\chi^2.$(i.e.$\chi^2$=$\chi^2$/ndof. )\\ 
Majorana neutrinos with 
($\nu_e\rightarrow\bar{\nu}_\mu$) have been used through this work. The  
results for Dirac neutrinos being practically identical.  

\section*{Acknowledgments}The authors are grateful to E. Akhmedov and D. Vignaud  
for their useful and informative comments. We are grateful to M.M. Guzzo and 
H. Nunokawa for useful discussions.  

  


\begin{table}[h]  
\begin{center}  
\begin{tabular}{ccccc}  
\hline  
\hline  
Experiment & Ref.& Data                   & Theory              & units \\  
\hline  
Homestake & \cite{exp1}  & $2.56 \pm 0.16 \pm 0.16$ &  $7.7^{+1.2}_{-1.0}$&  
SNU \\  
Kamiokande&\cite{exp5}  & 2.80$\pm$0.19$\pm$0.33 &  $5.15^{+1.0}_{-0.7}$&  
$10^6.cm^{-2}s^{-1}$ \\  
SAGE      &\cite{exp3}  & $66.6^{+7.8}_{-8.1}$ & $129^{+8}_{-6}$ & SNU \\  
GALLEX    &\cite{exp2}  & $77.5 \pm 6.2 ^{+4.3}_{-4.7}$ & $129^{+8}_{-6}$ & SNU \\  
SuperKamiokande & [8,9] & $2.44 \pm 0.05 ^{+0.09}_{-0.07}$ & $5.15^{+1.0}_{-0.7}$ &  
$10^6.cm^{-2}s^{-1}$ \\  
\hline  
\hline  
\end{tabular}  
\end{center}  
\caption{Neutrino event rates measured by solar neutrino  
experiments, and corresponding predictions from the BP98 solar model\cite{bp98}.  
The quoted errors are at $1\sigma$.}  
\label{data}  
\end{table}  


		\begin{figure}  
		\begin{center}  
		\mbox{  \epsfysize=16cm   
			\epsffile{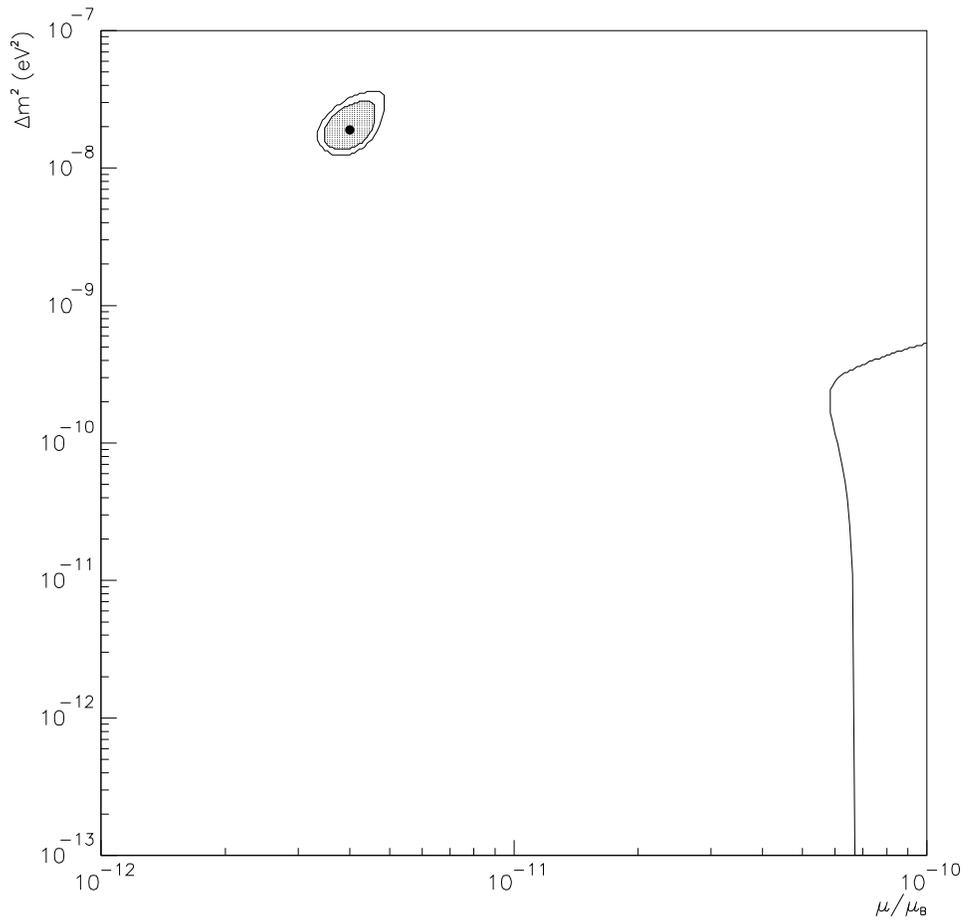}}  
		\end{center}  
                \caption{For the linear field we show the regions of parameters 
		$\Delta m^2$ and $\frac{\mu}{\mu_B}$, obtained by a fit to the
		total event rates only both at 95\%C.L.(doted erea) and 
		99\%C.L.(solide line). The best fit is indicated by dark filled
		circle.
		}  
		\label{1.ps}                  
		\end{figure}

		\begin{figure}  
		\begin{center}  
		\mbox{  \epsfysize=16cm   
			\epsffile{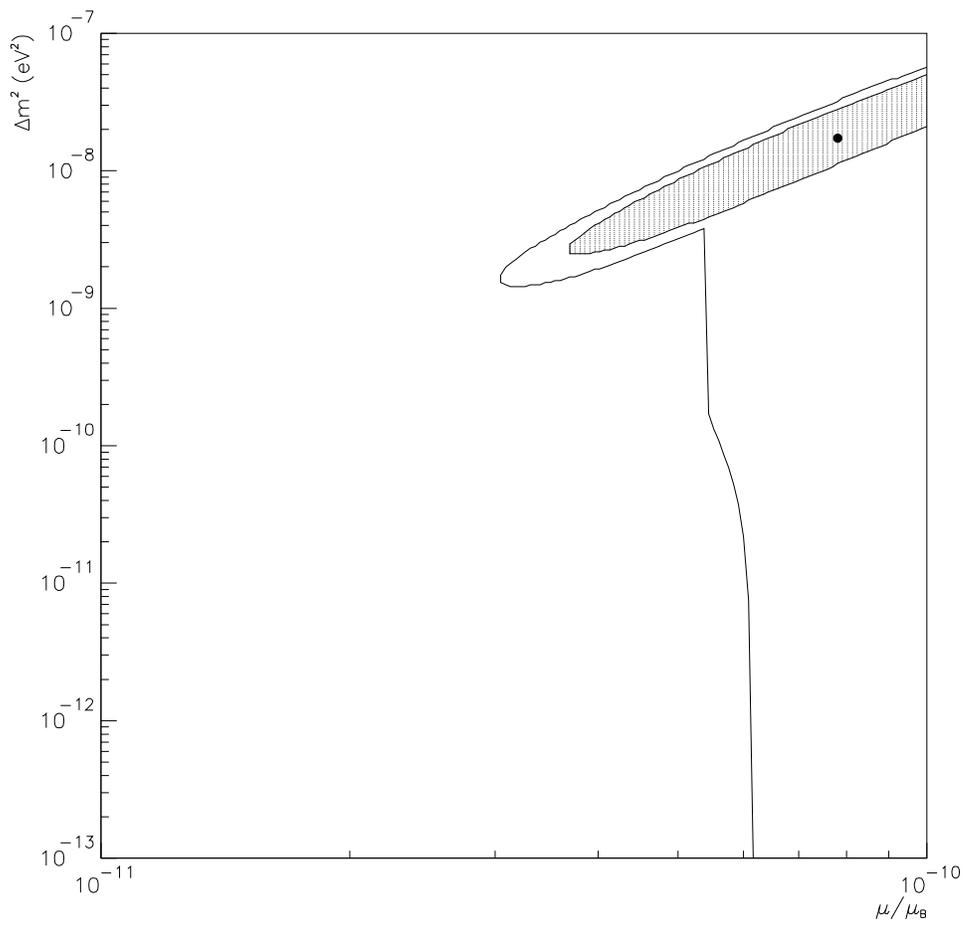}}  
		\end{center}  
 		\caption{Same as figure 1 using the Wood Saxon profile for the
		solar magnetic field.
		}
 		\label{2.ps}                  
		\end{figure}  

		\begin{figure}  
		\begin{center}  
		\mbox{  \epsfysize=16cm   
			\epsffile{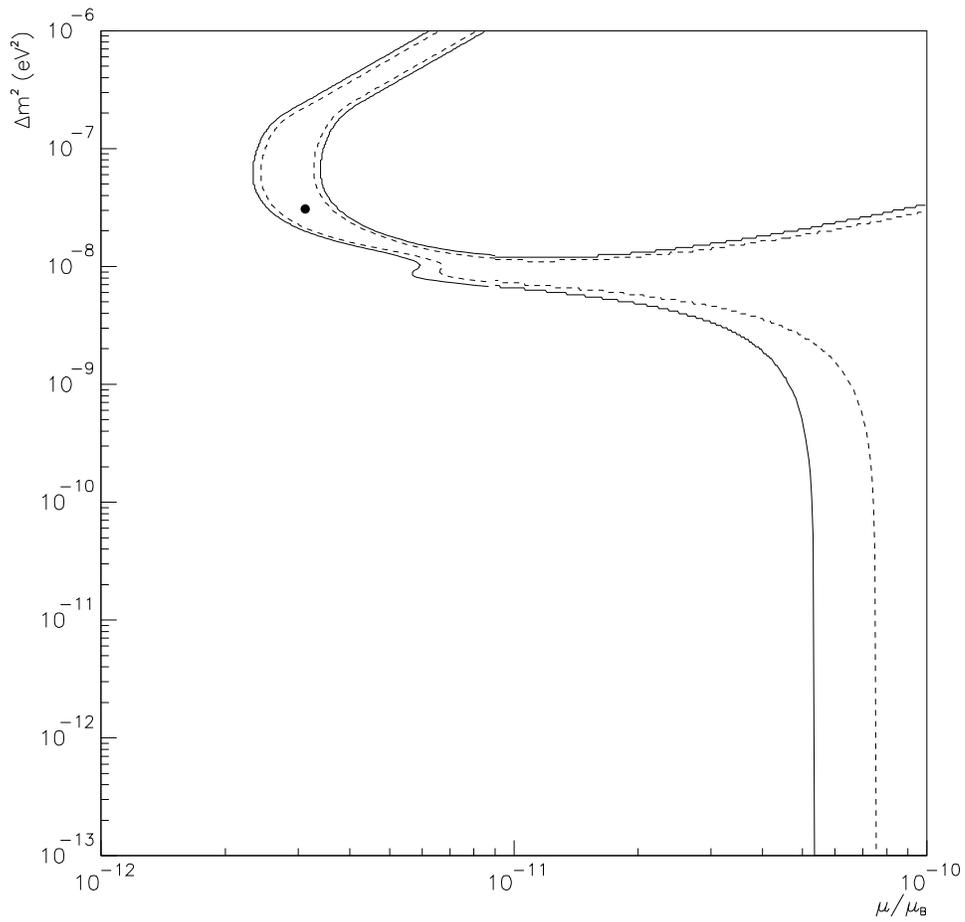}}  
		\end{center}  
		\caption{Considering only the SuperKamiokande energy spectrum,
		allowed region for neutrino parameters at 99\%C.L.(solide line)
		and 95\%C.L.(dashed line) are shown for the linear profile. The
		best fit is indicated bu a filled circle. 
		}
		\label{3.ps}                  
		\end{figure}  

		\begin{figure}  
		\begin{center}  
		\mbox{  \epsfysize=16cm   
			\epsffile{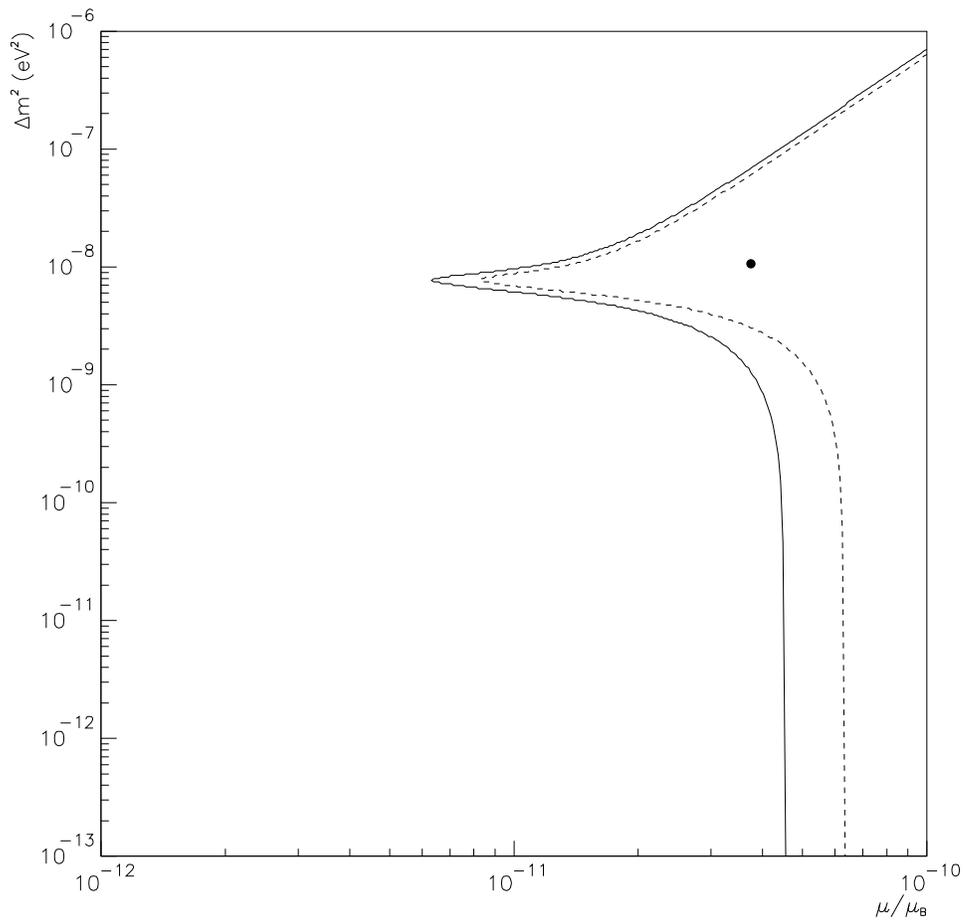}}  
		\end{center}  
		\caption{Same as figure 3 using the Wood Saxon field.
		} 	
		\label{4.ps}                  
		\end{figure}

		\begin{figure}  
		\begin{center}  
		\mbox{  \epsfysize=16cm   
			\epsffile{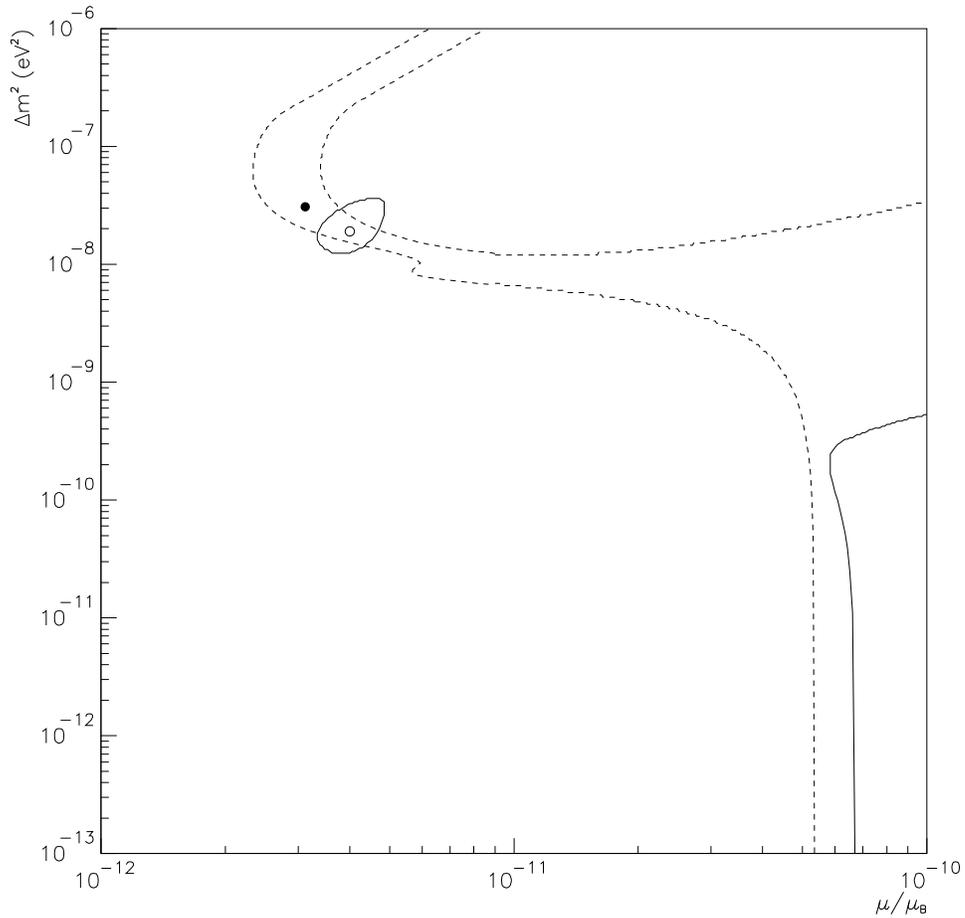}}  
		\end{center}  
  		\caption{For the linear field(eq.8) we compare the allowed neutrino
		parameters regions obtained separately by the total event rates (solide
		line) and the SuperKamiokande energy spectrum (dashed line). The
		filled circle is the best fit to the SuperKamiokande spectrum
		(eq.12) while the open circle is the best fit to the total event
		rates(eq.10). The comparaison is done at 99\%C.L..
		}
  		\label{5.ps}                  
		\end{figure}

		\begin{figure}  
		\begin{center}  
		\mbox{  \epsfysize=16cm   
			\epsffile{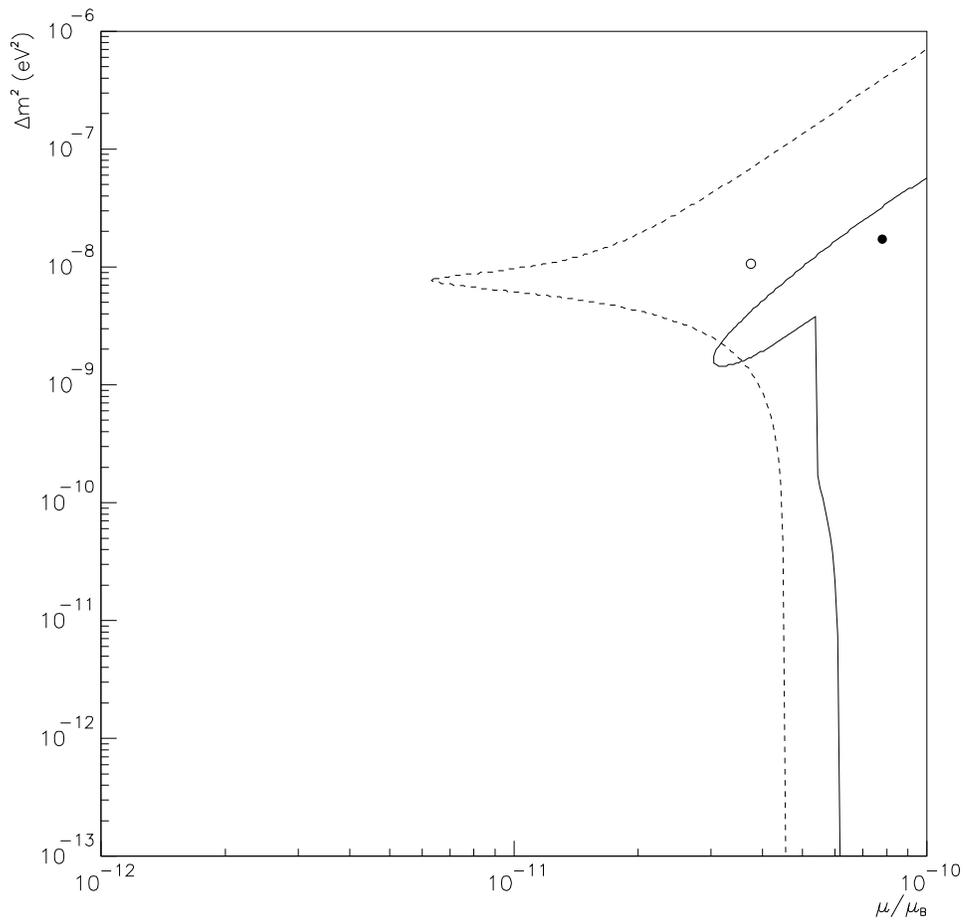}}  
		\end{center}  
		\caption{Same as in figure 5 for the Wood Saxon filed.
		}
		\label{6.ps}                  
		\end{figure}

		\begin{figure}  
		\begin{center}  
		\mbox{  \epsfysize=16cm   
			\epsffile{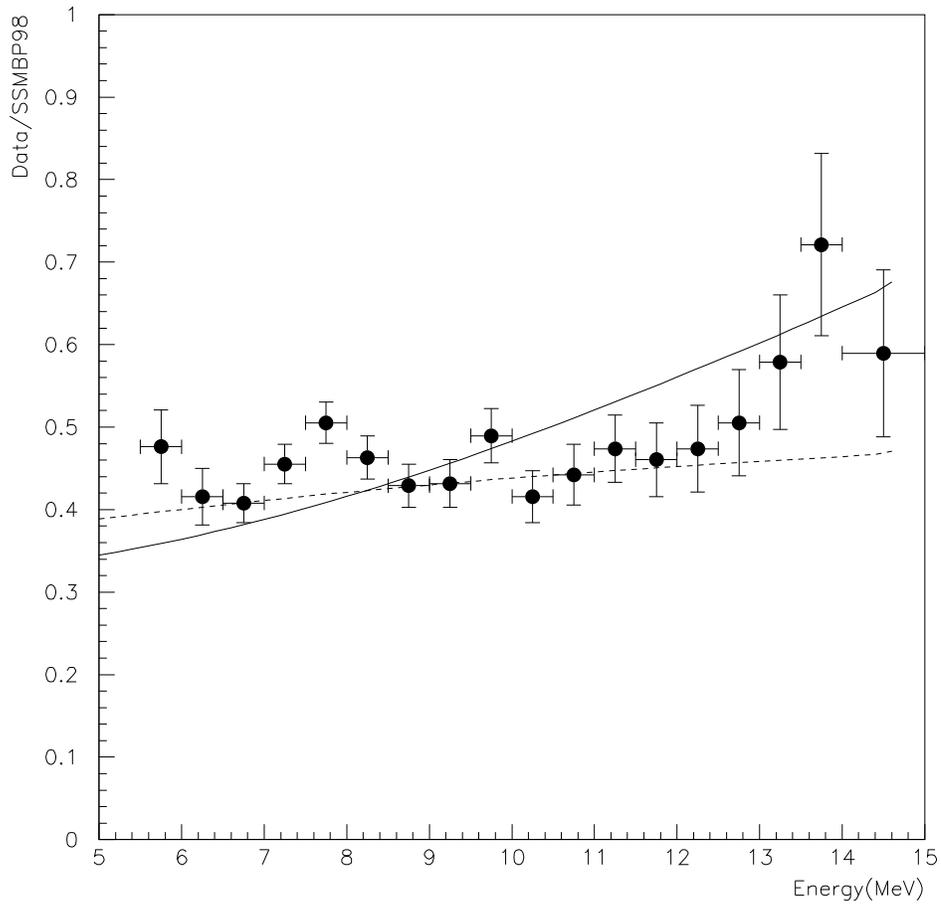}}  
		\end{center}  
		\caption{We plot the recoil energy spectra expected from RSFP
		scenario using the best fit parameters(eqs.10-11) for the linear
		field (solide line) and Wood Saxon filed (doted line) divided by
		the SSM expectation. SuperKamiokande data are also shown with error
		bars representing the statistical and systematical errors added in
		quadrature. The data were directly read from M.B. Smy in
		\cite{exp4}.
		}
		\label{7.ps}                  
		\end{figure}

		\begin{figure}  
		\begin{center}  
		\mbox{  \epsfysize=16cm   
			\epsffile{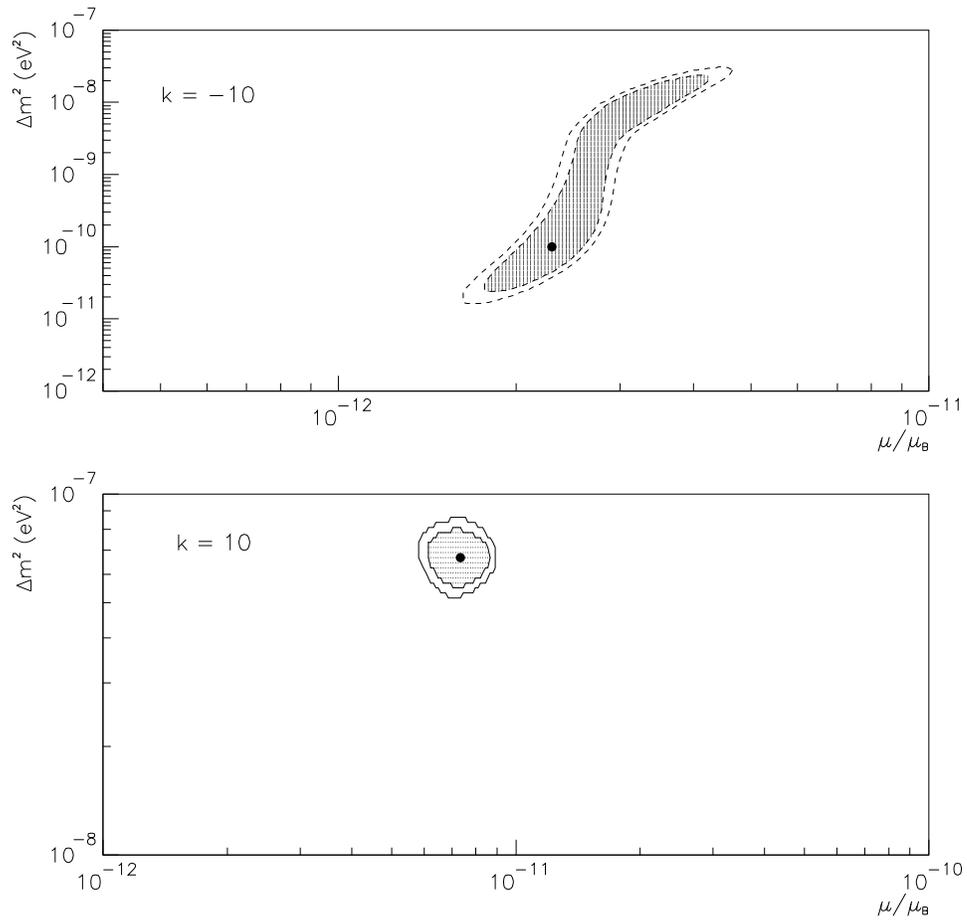}}  
		\end{center}  
 		\caption{Regions of the parameters $\Delta m^2$ and
		$\frac{\mu}{\mu_B}$, obtained by a fit to the total event rates
		only, using the field distribution given by eq.8 and the two values
		of k(see the text for definition). The best fit is indicated by
		dark circles, results are at 99\%C.L. and 95\%C.L.(doted erea).
		}
 		\label{8.ps}                  
		\end{figure}

		\begin{figure}  
		\begin{center}  
		\mbox{  \epsfysize=16cm   
			\epsffile{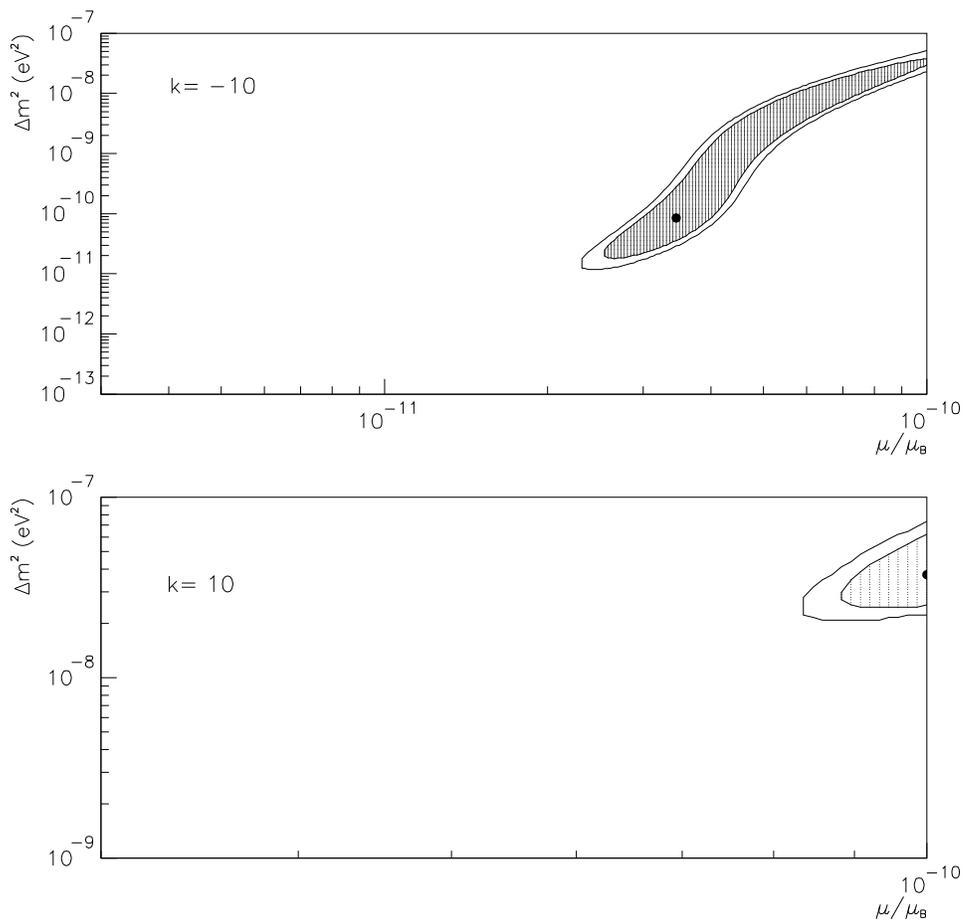}}  
		\end{center}  
		\caption{Same as in figure 8 for the solar field given by eq.9}
		\label{9.ps}                  
		\end{figure}

		\begin{figure}  
		\begin{center}  
		\mbox{  \epsfysize=16cm   
			\epsffile{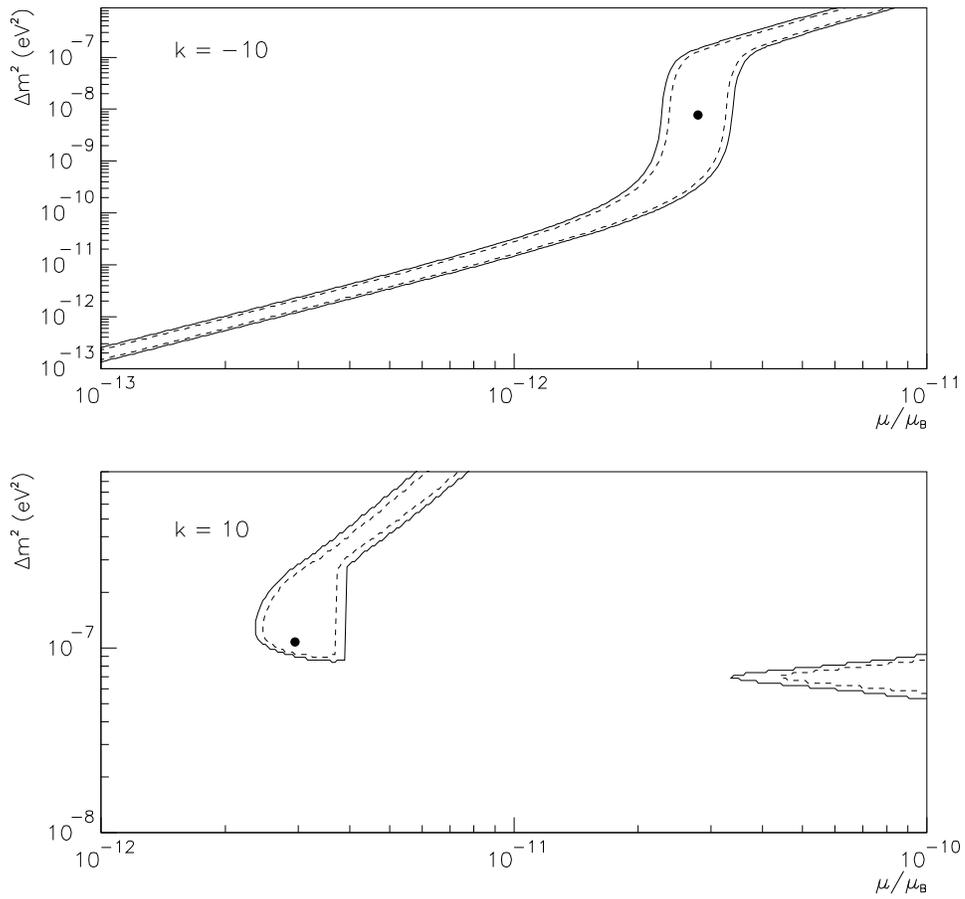}}  
		\end{center}  
 		\caption{Considering only the SuperKamiokande data spectrum, the
		allowed regions for neutrino parameters at 99\%C.L. (solide line)
		and 95\%C.L.(dashed line) are shown for the linear profile and two
		values of k.
		}
 		\label{10.ps}                  
		\end{figure}  

		\begin{figure}  
		\begin{center}  
		\mbox{  \epsfysize=16cm   
			\epsffile{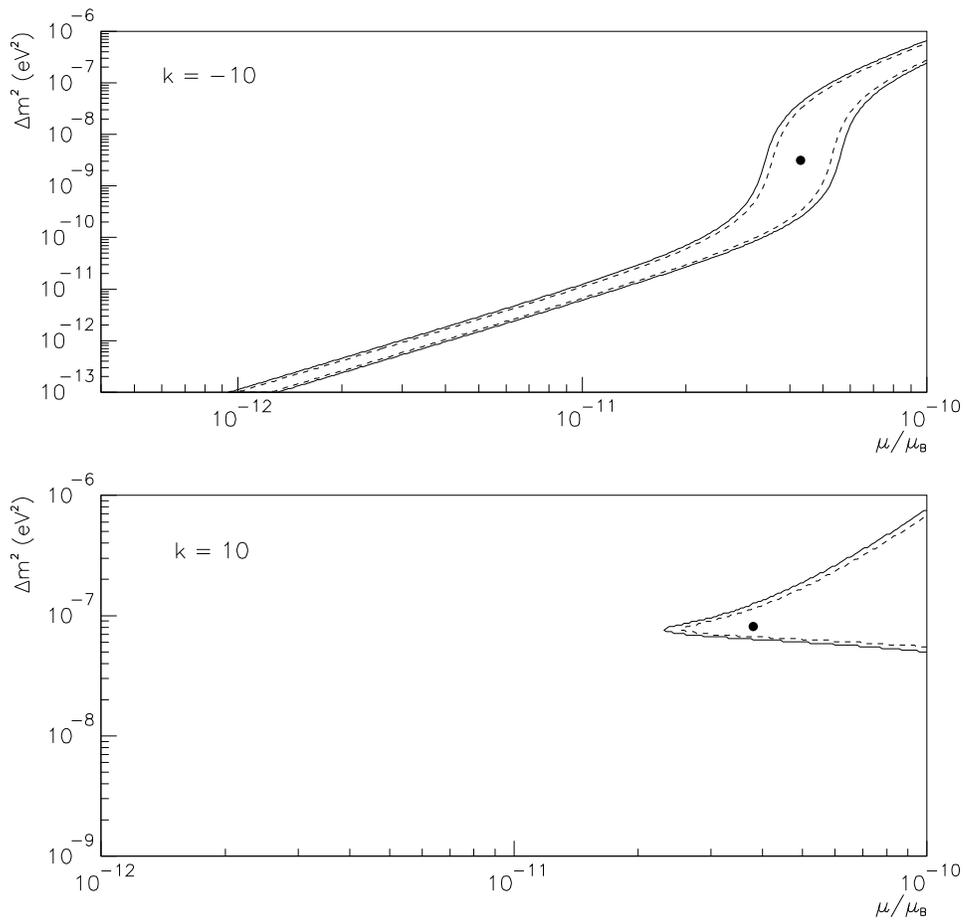}}  
		\end{center}  
		\caption{Same as in figure 10 using the Wood Saxon field.
		}
		\label{11.ps}                  
		\end{figure}  

		\begin{figure}  
		\begin{center}  
		\mbox{  \epsfysize=16cm   
			\epsffile{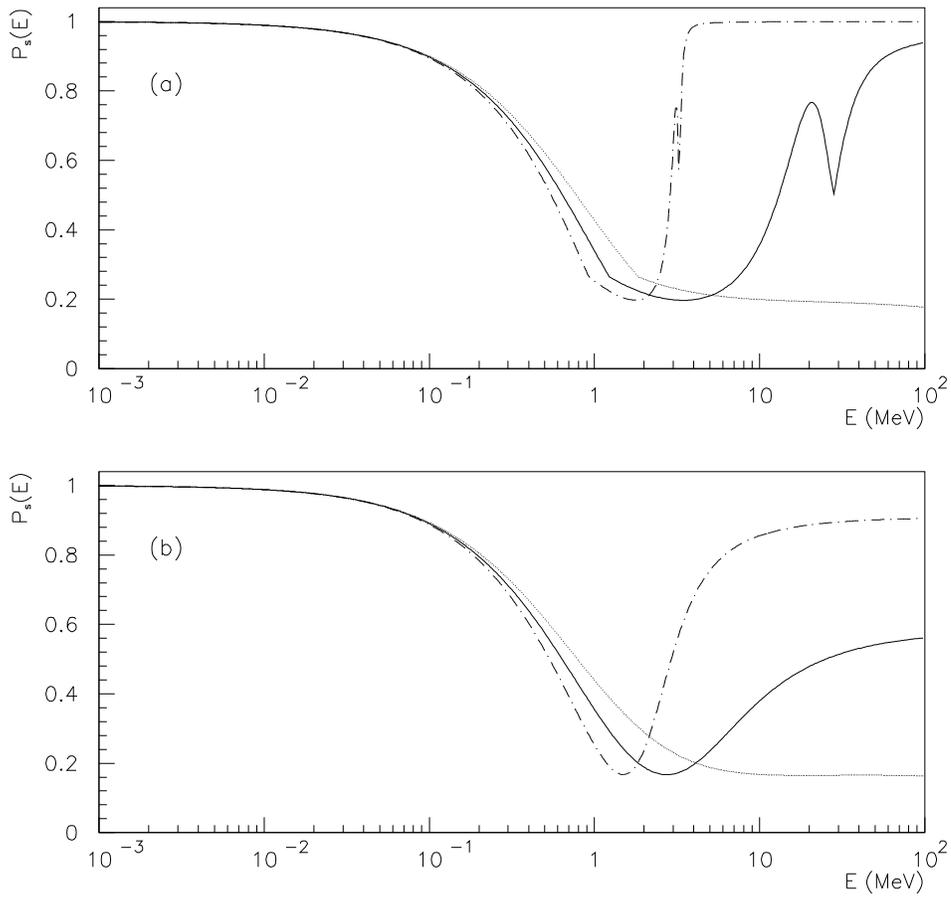}}  
		\end{center}  
                \caption{Neutrino survival probability computed using the  
		linear(a) and the Wood Saxon(b). The neutrino parameters are  
		those of the static best fit found in eqs.10-11. The solid lines  
		correspond to the static case. Doted and dash-doted lines refer  
		to the rotated case respectively with k = -10 and k = +10.}  
		\label{12.ps}                  
		\end{figure}  

		\begin{figure}  
		\begin{center}  
		\mbox{  \epsfysize=16cm   
			\epsffile{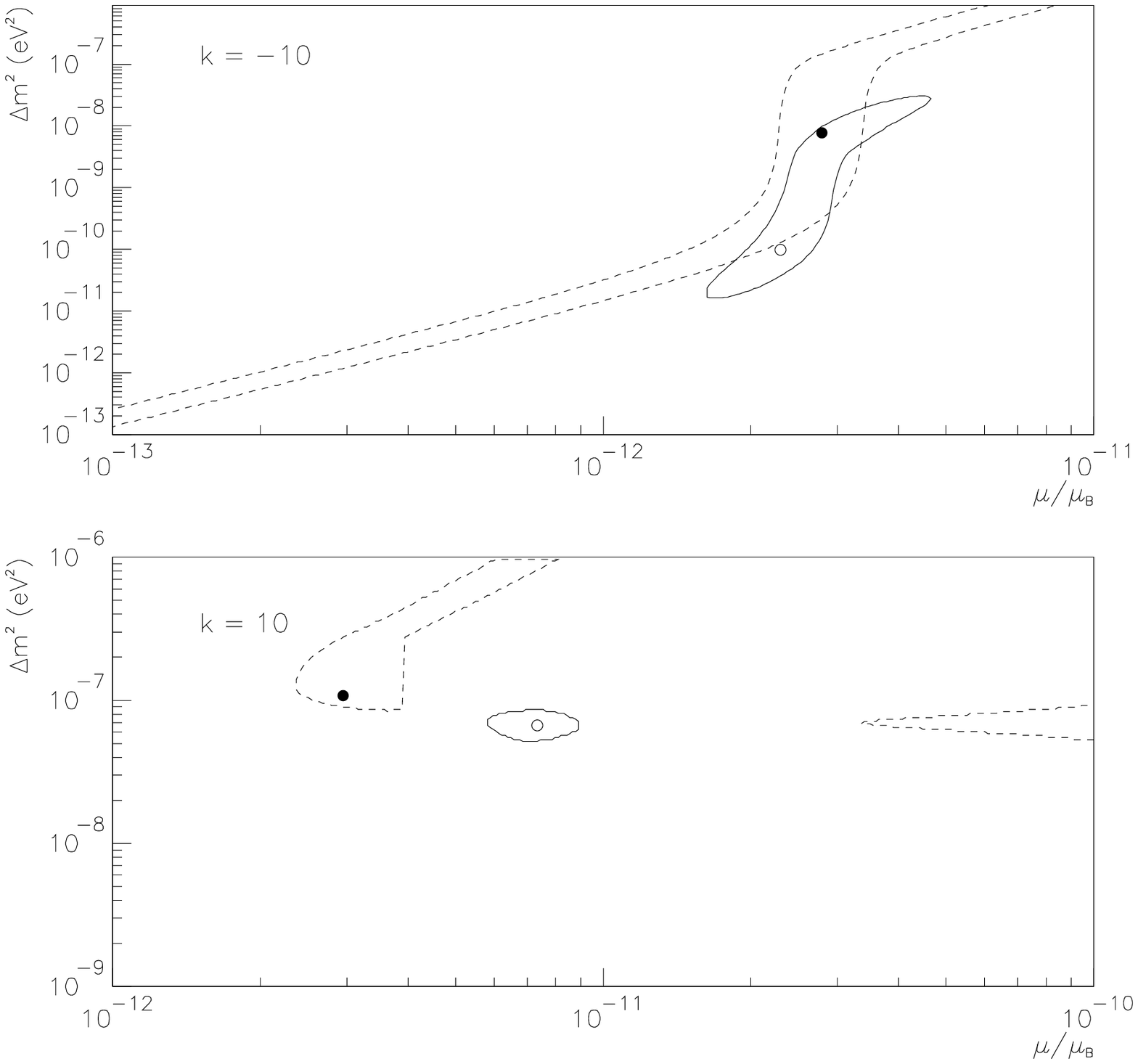}}  
		\end{center}  
		\caption{For the linear field(eq.8) and the two values of k, we
		compare the allowed neutrino parameters regions obtained by
		considering only constraints coming from the total event rates(figure 8) 
		with those coming from the SuperKamiokande data spectrum 
		(figure 10).
		}
		\label{13.ps}                  
		\end{figure}  

		\begin{figure}  
		\begin{center}  
		\mbox{  \epsfysize=16cm   
			\epsffile{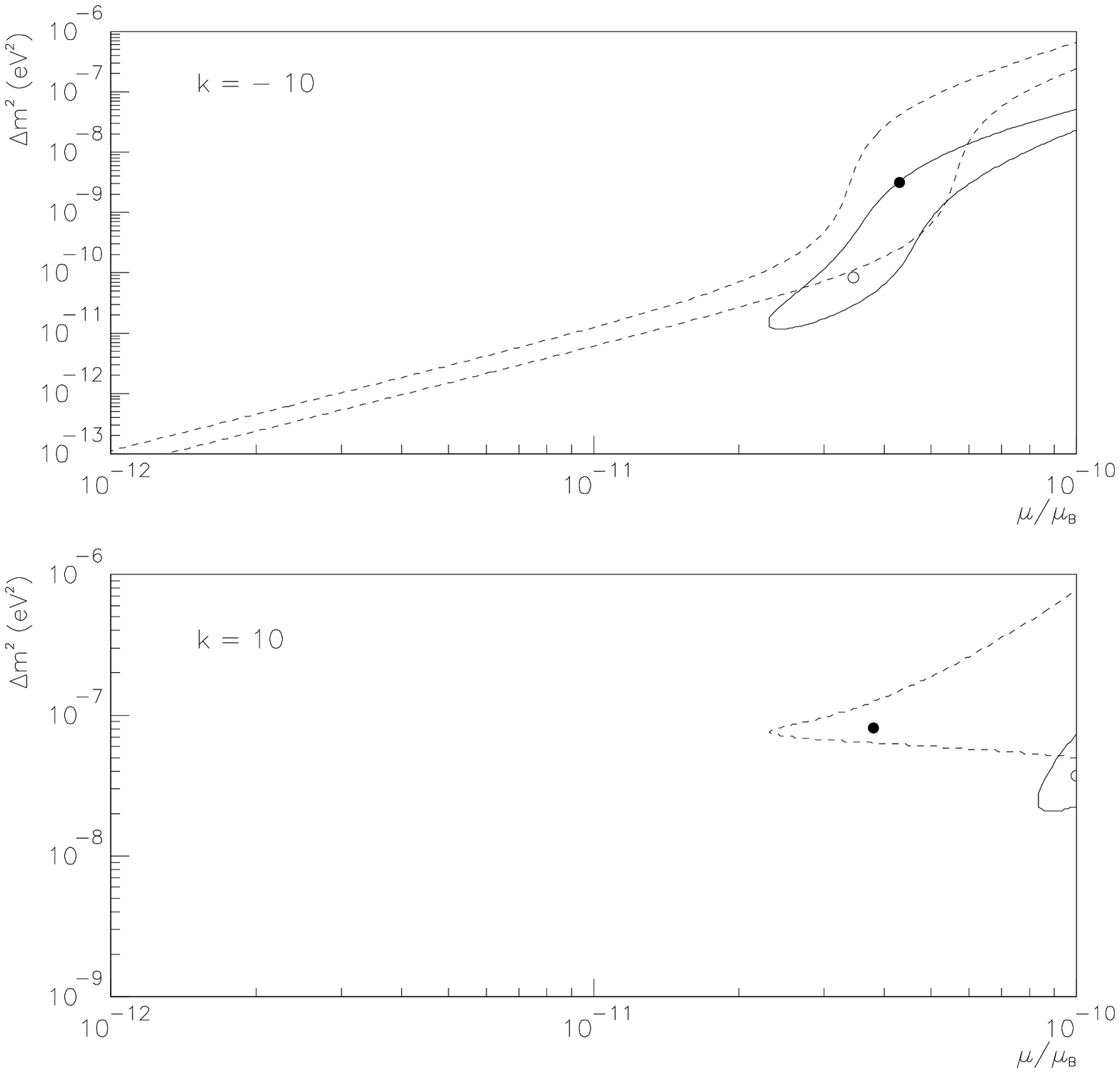}}  
		\end{center}  
		\caption{For the Wood Saxon field (eq.9) and the two values of k,
		we compare the allowed neutrino parameters regions obtained by
		considering only constraints coming from the total event rates
		(figure 9) with those coming from the SuperKamiokande data spectrum
		(figure 11).
		}
		\label{14.ps}                  
		\end{figure}

		\begin{figure}  
		\begin{center}  
		\mbox{  \epsfysize=16cm   
			\epsffile{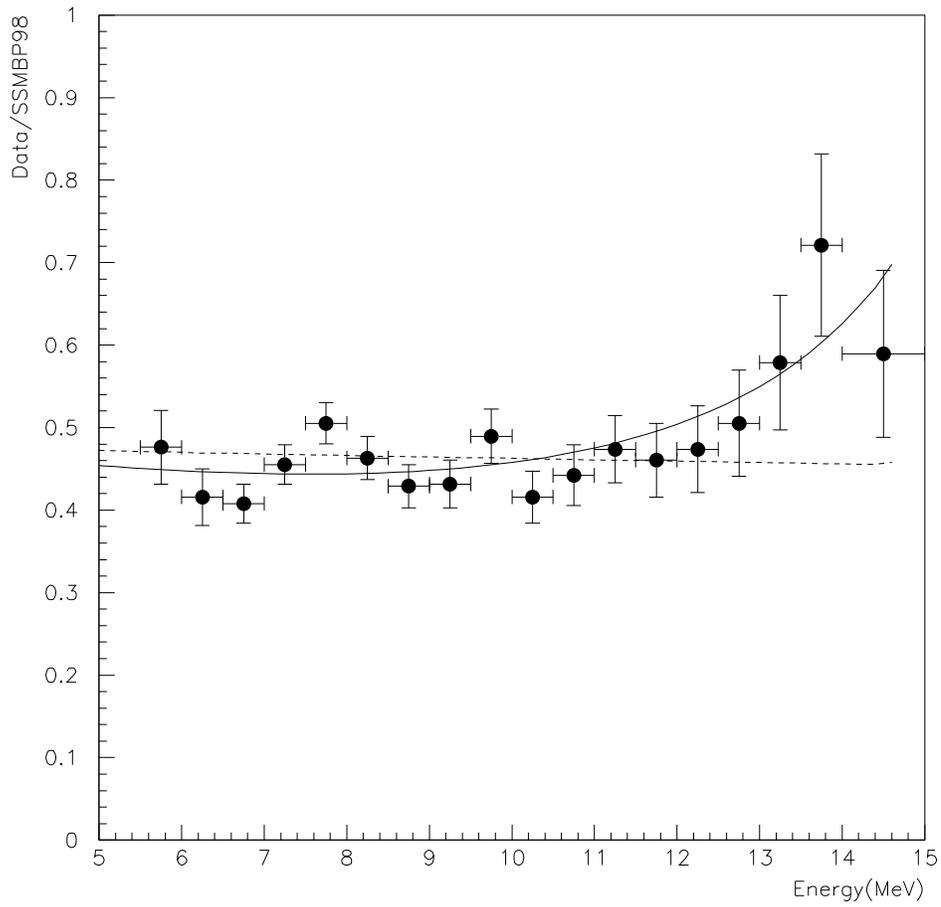}}  
		\end{center}  
		\caption{As in figure 7, for the linear field (solide line)  and
		Wood Saxon field (doted line) and for k=-10, we plot the recoil
		electron energy spectra from RSFP scenario using the best fit
		parameters(eq.21 and eq.22) divided by the SSM prediction.
		}
		\label{15.ps}                  
		\end{figure}  
\end{document}